\documentclass[twocolumn,preprintnumbers,secnumarabic,amsmath,amssymb,superscriptaddress,nofootinbib]{revtex4}

\usepackage{graphicx}
\usepackage{dcolumn}
\usepackage{bm}

\def\be{\begin{equation}}
\def\ee{\end{equation}}
\def\bea{\begin{eqnarray}}
\def\eea{\end{eqnarray}}

\newcommand{\0}{\textrm{\tiny{(0)}}}
\newcommand{\1}{\textrm{\tiny{(1)}}}

\newcommand{\2}{\textrm{\tiny{(2)}}}

\begin{document}


\title{(In)finite extensions of algebras from their \.{I}n\"on\"u-Wigner contractions}

\medskip\
\author{Oleg Khasanov}
\affiliation{{School of Mathematical Sciences\\
  The Raymond and Beverly Sackler Faculty of Exact Sciences\\
  Tel Aviv University, Ramat Aviv, 69978, Israel\\
  \textsf{oleg@post.tau.ac.il}
}}

\author{Stanislav Kuperstein}
\affiliation{LPTHE, Universit\'e Pierre et Marie Curie -- Paris 6\\
CNRS UMR 7589, 4 Place Jussieu, 75252 Paris Cedex 05, France\\
\textsf{skuperst@lpthe.jussieu.fr}}


\begin{abstract}

The way to obtain massive non-relativistic states from the Poincar\'e algebra is twofold. First,
following \.{I}n\"on\"u and Wigner the Poincar\'e algebra has to be contracted to the Galilean one. Second,
the Galilean algebra is to be extended to include the central mass operator. We show that
the central extension might be properly encoded in the non-relativistic contraction.
In fact, any \.{I}n\"on\"u-Wigner contraction of one algebra to another,
corresponds to an infinite tower of abelian extensions of the latter.
The proposed method is straightforward and
holds for both central and non-central extensions.
Apart from the Bargmann (non-zero mass) extension of the Galilean algebra,
our list of examples includes the Weyl algebra obtained from an extension of
the contracted $SO(3)$ algebra,  the Carrollian (ultra-relativistic) contraction of the Poincar\'e algebra, the exotic
Newton-Hooke algebra and some others.

\center{
    \emph{The paper is dedicated to the memory of Laurent Houart (1967-2011).}
}

\end{abstract}

\maketitle

\section{\bf Introduction}

The \emph{\.{I}n\"on\"u-Wigner contraction} (IW) \cite{Inonu:1953sp} (see \cite{WeimarWoods} for the generalizations) plays a very important r\^ole in physics.
To see the most basic example of the contraction, start with the $SO(3)$ algebra of $3d$ rotations. The algebra
consists of three operators $J_i$ satisfying the commutator relations:
\be
[J_2, J_3] = i J_1 \, , [J_2, J_3] = i J_1 \quad \textrm{and} \quad [J_1, J_2] = i J_3 \, .
\ee
Rescaling the operators $J_1$ and $J_2$:
\be  \label{WeylRescaling}
J_{i=1,2} = \tilde{J}_{i=1,2}/\sigma  \quad J_{3} = \tilde{J}_{3}
\ee
and taking the $\sigma \to 0$ limit one ends up with a new algebra:
\be
\label{tildeJJJ}
[\tilde{J}_2, \tilde{J}_3] = i \tilde{J}_1 \, , [\tilde{J}_3, \tilde{J}_1] = i \tilde{J}_2
\quad \textrm{and} \quad [\tilde{J}_1, \tilde{J}_2] = 0 \, ,
\ee
which is the algebra of $ISO(2)$, the group of the $\mathbb{R}^2$ isometries, where $\tilde{J}_{1,2}$
and $\tilde{J}_{3}$ are the translations and the rotation respectively. One can think of this contraction
as of taking to infinity the radius of the 2-sphere, while zooming near a point on it.

Probably the most famous physical application of the IW contraction
is the \emph{non-relativistic} limit of the Poincar\'e algebra. As we will briefly review
in the paper, the contraction parameter is the speed of light $c$ and the rescaled operators are
the time translation (the Hamiltonian) and the boosts. By the construction, however, the $c \to \infty$
limit leads to the Galilean algebra, which has exactly the same number of generators
as the original Poincar\'e algebra. For physical applications, though, it is necessarily to \emph{centrally
extend} the non-relativistic algebra by including the mass parameter $M$. The $M \neq 0$ version of the
Galilean algebra is commonly known as the Bargmann algebra \cite{Bargmann:1954gh}.

In fact, the Poincar\'e algebra is itself a contraction of the de Sitter algebra $SO(4,1)$,
where the contraction parameter now is
the cosmological constant $\Lambda$.
As was shown \cite{Bacry:1968zf} by Bacry and L\'evy-Leblond in 1967, under some physical assumptions (like parity,
time-reversal, \emph{etc.}) the full list of possible kinematic groups consists of the (anti) de Sitter groups and their
IW contractions.  For example, one of the possibilities is the $(c, \Lambda) \to (\infty,0)$ contraction
of the de Sitter algebra with the parameter $\omega = c \sqrt{\Lambda}$ kept fixed.
The contraction produces the so-called Newton-Hooke (NH) algebra,
which further reduces to the Galilean algebra for $\omega=0$.
One can also take the \emph{ultra-relativistic} ($c \to 0$) limit of the Poincar\'e algebra arriving at the Carrollian algebra \cite{Levy-Leblond1965,SenGupta1966}, the much less studied counterpart of the Galilean algebra.

In quantum mechanics we are interested in general in \emph{projective} representations of groups.
On the other hand, a projective representation of a group $G$ is essentially equivalent to
a regular representation of the central extension of $G$. For this reason
central extensions are of special importance in quantum physics. The canonical (and the simplest)
example is the Weyl algebra:
\be
\left[q, p\right] = i \hbar \, .
\ee
Unlike the Galilean group, both the Poincar\'e algebra has no central extensions
except for the two dimensional case.\footnote{
The Galilean and the NH algebra have an additional, the so called \emph{exotic},
central extension in $2+1$ dimensions \cite{LevyLeblond:1971} as we will describe in the paper.}
This can be seen from the group cohomology arguments (see \cite{deAzcarraga:1995jw} for a detailed review).
These groups, although, still have plenty of \emph{non-central}
extensions, some of which are interesting from the physical point of view, \emph{e.g.}
the Maxwell algebra \cite{Bacry:1970ye,Schrader:1972zd}, a specific non-central extension of the Poincar\'e algebra.

The main goal of this paper is to show that starting from an IW contraction one can straightforwardly
find an extension of the contracted algebra.
Although the proposed procedure explicitly yields an infinite extension, it can be easily truncated at any level.
It works both for central and non-central extensions, and we find a criterion for an extension to be central.

Our first example is a IW contraction of $SO(3)$
given by the rescaling:
\be\
\label{SO3scaling}
J_{1,2} = \tilde{J}_{1,2}/\sigma^2 \quad J_{3} = \tilde{J}_{3}/\sigma^3 \, .
\ee
For $\sigma \to 0$ we find an abelian algebra (all the commutators vanish), but as we demonstrate in the paper
the first level extension dictated by this contraction reproduces the central extension of the Weyl algebra.
The higher level extensions, however, are not anymore central.


The organization of the paper is as follows. In the next section we present our method for constructing
algebra extensions from its \.{I}n\"on\"u-Wigner contraction. We then describe various applications,
like the non-central extension of the Poincar\'e algebra from the de Sitter algebra contraction,
the Galilean and the Carrollian contractions of the Poincar\'e algebra and the exotic extension of the Galilean and the NH algebra.
We end up with a short list of open questions.

\section{\bf The method}

\subsection{Lie algebra bundles}

For our construction we will need the notion of bundles of Lie algebras \cite{LieAlgebraBundle} (see also \cite{Weinstein} for a pedagogical introduction).
A \emph{Lie algebra bundle} (Lie bundle for short) is a vector bundle for which each fiber
has a smoothly varying Lie algebra structure.\footnote{
We will deal exclusively with Lie algebras and not with Lie groups.
Let us only notice here that every bundle of Lie groups defines a bundle of Lie algebras, and
every bundle of algebras can be integrated to a bundle of groups.
The group bundle, however, is not necessarily Hausdorf (see \cite{Weinstein}).}
More explicitly, the vector bundle $(\mathcal{E}, \pi, \mathcal{S})$ should be equipped with a morphism
$\theta : \, \mathcal{E} \otimes \mathcal{E} \to \mathcal{E}$, which induces a Lie algebra structure on
each fiber $\mathcal{E}_\sigma$. Here $\mathcal{E}$ is the total space, $\pi$ is the projection map and $\sigma$ is
a point on the base $\mathcal{S}$. This definition is sometimes called in the literature a \emph{weak}
Lie algebra bundle, in contrast to the \emph{strong} one which requires also
local triviality of the Lie structure.
It means that for any $\sigma \in \mathcal{S}$ there exists a neighborhood $U_\sigma$ of $\sigma$, a Lie algebra $g$
and a morphism $\phi : \, g  \times U_\sigma \to \pi^{-1} (U_\sigma)$ such that
$\left. \phi \right\vert_{\sigma^\prime} : \, g  \to \pi^{-1} (\sigma^\prime)$
is a Lie algebra isomorphism for each $\sigma^\prime \in U_\sigma$.
We will refer to $\phi$ as a Lie bundle trivialization.
If $\phi$ extends to the entire base, the Lie algebra bundle will be called \emph{trivial}.

Next, let us restrict our attention to weak Lie bundles, where the base $\mathcal{S}$ is just an affine
line\footnote{Our discussion can be easily generalized to complex numbers.} $\mathbb{R}^1$.
In this case any Lie algebra bundle will be trivial as a vector bundle, but not
necessarily so from the Lie algebra point of view. The Lie algebra $g$ at any fiber is the same
as a vector space, but its brackets are in general $\sigma$-dependent.
If $a^i$ are the generators of $g$ and $a_i^\sigma=(a_i,\sigma)$ denotes a point on the Lie bundle,
then the most general form of the commutators
is:
\be
\left[ a_i^\sigma, a_j^\sigma \right] = f^k_{ij} (\sigma) a_k^\sigma \, ,
\ee
where $f^k_{ij} (\sigma)$ are some smooth functions.

Let us consider the following example.
Assume that $i$ runs from $1$ to $3$ and:
\be
\label{f-ijk}
f^k_{ij} (\sigma) = \frac{1}{2} \epsilon^k_{ij} \cdot f(\sigma) \, .
\ee
If $f(\sigma) \neq 0$ for any $\sigma \in \mathbb{R}^1$, then the Lie bundle is trivial
(namely strong) with the trivialization $\phi$ given by:
\be
\label{fsigma}
\phi(a_i,\sigma)= \frac{a_i^\sigma}{f(\sigma)}  \,
\ee
and $a_i$'s identified with the $SO(3)$ generators $J_i$'s.
On the other hand, the Lie bundle will be non-trivial\footnote{If the base space is $\mathbb{R}^1$ local triviality implies also global triviality.} (weak but not strong) if $f(\sigma_0)= 0$ for some
$\sigma_0 \in \mathbb{R}^1$, since in this case $\phi$ is ill-defined for $\sigma=\sigma_0$.
We still, however, can claim that the Lie bundle is trivial everywhere \emph{except} at
$\sigma=\sigma_0$.

Similarly we can see the rescaling (\ref{WeylRescaling}) as a trivialization map for the Lie algebra bundle given by:
\be
\label{tildeJJJ2}
[\tilde{J}_2, \tilde{J}_3] = i \tilde{J}_1 \, , [\tilde{J}_3, \tilde{J}_1] = i \tilde{J}_2
\quad \textrm{and} \quad [\tilde{J}_1, \tilde{J}_2] = \sigma^2 \tilde{J}_3 \, .
\ee
At $\sigma=0$ the trivialization map is singular and the Lie algebra structure reduces to (\ref{tildeJJJ}), but at any other point the algebra is isomorphic to $SO(3)$.

\subsection{The \.{I}n\"on\"u-Wigner bundle}

We are now in a position to define the main ingredient of our construction.

{\bf Definition.}
\emph{
The \.{I}n\"on\"u-Wigner (IW) bundle of a Lie algebra $g$ is a (weak) Lie algebra bundle over an affine line $\mathbb{R}^1$,
such that its restriction to $\mathbb{R}^1 \! \setminus \! \{0\}$ is a trivial bundle $g \times \mathbb{R}^1 \! \setminus \! \{0\}$.
}

From the above discussion the Lie bundle defined by (\ref{tildeJJJ2}) and (\ref{f-ijk}) are examples of IW bundles. The latter is non-trivial if and only if $f(0)=0$.

As yet another example consider the following IW Lie algebra bundle\footnote{
The form of the commutators in (\ref{Weyl1}) suggests an analogy with \emph{loop algebras}. The similarity is, however, illusive,
since the parameter $\sigma$ takes values in $\mathbb{R}^1$ and not in $\mathbb{S}^1$, and, even more importantly, for the loop algebra construction, the Lie algebra necessarily remains the same all over the loop, while in our case, a new algebra emerges at
$\sigma=0$.
}:
\be
\label{Weyl1}
[J_2^\sigma, J_3^\sigma] = i \sigma^3 J_1^\sigma  , \, [J_3^\sigma, J^\sigma_1] = i \sigma^3 J^\sigma_2  ,
\, [J_1^\sigma, J_2^\sigma] = i \sigma J_3^\sigma \, .
\ee
Upon using the trivialization map:
\be
\label{Weyl2}
\phi(J_{1,2}, \sigma) =\frac{J_{1,2}^\sigma}{\sigma^2} \quad \textrm{and} \quad \phi(J_{3}, \sigma) =\frac{J_{3}^\sigma}{\sigma^3} \, .
\ee
we find that for $\sigma \neq 0$ the algebra is isomorphic to $SO(3)$. We see that
the trivialization morphism $\phi$ is actually the rescaling of the generators in (\ref{SO3scaling}).
The fact that $\phi$ fails to be an isomorphism at $\sigma=0$ indicates that
we find a new algebra at this point.

\subsection{The extensions}

Our goal in this subsection is to explore the $\sigma \to 0$ limit.

Let $\widetilde{\mathfrak{g}}$ stand for a vector space of
\emph{smooth} sections of an IW bundle of an algebra $g$. Again, as a vector space $g$
is the same all over the base (and so we drop here the $\sigma$-index),
but its commutators are not. The (infinite) vector space of smooth sections $\widetilde{\mathfrak{g}}$
is spanned 
by the vectors $\sigma^n a_i$, where $a_i \in g$ and $n \geqslant 0$.
Moreover, this vector space is a
Lie algebra by its own right, since we can multiply the sections pointwise.
Sections vanishing at $\sigma=0$ is an ideal of this algebra. This ideal is $\sigma \widetilde{\mathfrak{g}}$.

The \emph{classical IW algebra} is the quotient:
\be
\mathfrak{g}_0 \equiv \widetilde{\mathfrak{g}}/ {\sigma \widetilde{\mathfrak{g}}} \, .
\ee
To arrive at this algebra it is enough to directly take the $\sigma \to 0$ limit.
Notice that $\mathfrak{g}_0$ and $g$ are identical as vector spaces, but not as Lie algebras.
For (\ref{Weyl1})
the algebra $\mathfrak{g}_0$ is an abelian algebra of $J_1, J_2$ and $J_3$.
Similarly, the scaling (\ref{tildeJJJ}) leads to the $ISO(2)$ algebra.

Up to this point, the $\sigma \to 0$ contraction is identical to the original \.{I}n\"on\"u and Wigner prescription.
We, however, don't want to stop here. Notice that $\sigma^n \widetilde{\mathfrak{g}}$ is an ideal of $\widetilde{\mathfrak{g}}$
for any $n \geqslant 1$. The \emph{semi-classical IW algebra} is defined by:
\be
\mathfrak{g}_1 \equiv \widetilde{\mathfrak{g}}/ {\sigma^2 \widetilde{\mathfrak{g}}} \, .
\ee
The algebra $\mathfrak{g}_1$ has \emph{twice} more generators than $\mathfrak{g}_0$, but it might happen that some
of these new generators form an ideal and can be quotiented out.
For the IW bundle defined by
(\ref{Weyl1}) let us introduce $J_i^{(0)} \equiv J_i$ and $J_i^{(1)} \equiv \sigma J_i$.
We find that the semi-classical algebra $\mathfrak{g}_1$ has only one non-trivial commutator:
\be
\label{WeylJ}
\left[ J_1^{(0)}, J_2^{(0)} \right] = i J_3^{(1)} \, .
\ee
As was announced in Introduction this is the central extension of the Weyl algebra.
To be more precise, in order to identify  $\mathfrak{g}_1$ with the Weyl algebra
we have to quotient the algebra by its abelian ideal $\left\{ J_3^{(0)},J_1^{(1)},J_2^{(1)} \right\}$.

We can easily repeat this procedure for any $n$. Clearly, the new algebra will have $n$ times more generators than the original algebra. For the IW bundle in (\ref{Weyl1}) the
$n$ \emph{level} algebra $\mathfrak{g}_n \equiv \widetilde{\mathfrak{g}}/ {\sigma^{n+1} \widetilde{\mathfrak{g}}}$ is
(here $J_i^{(n)}$ stands for $\sigma^n J_i$ and $J_i^{(n)}=0$ for negative $n$):
\bea
&&  \left[ J_2^{(n)}, J_3^{(m)} \right] = i J_1^{(n+m-3)}, \, \left[ J_3^{(n)}, J_1^{(m)} \right] = i J_2^{(n+m-3)}  , \nonumber \\
&& \qquad  \left[ J_1^{(n)}, J_2^{(m)} \right] = i J_3^{(n+m-1)}
\eea
with all the other commutators vanishing. One may argue that the output algebra is not particularly interesting.
The situation changes drastically, however, for a more sophisticated trivialization (rescaling) $\phi$
than the one in (\ref{Weyl2}).

Before concluding this section, let us discuss how our semi-classical extension $\mathfrak{g}_1$
may lead to a \emph{central} extension of the $\mathfrak{g}_0$ algebra. To this end it is worth recalling the
rigorous definition of a (not necessarily central) extension. An algebra $\mathfrak{a}^\prime$ is called an extension of
$\mathfrak{a}$ by an ideal $\mathfrak{n}$, if $\mathfrak{a}$ is isomorphic to the quotient $\mathfrak{a}^\prime/\mathfrak{n}$.
This definition can be shortly summarized with the following \emph{short exact sequence}:
\be
0 \to \mathfrak{n} \to \mathfrak{a}^\prime\to \mathfrak{a} \to 0 \, .
\ee
If $\mathfrak{n}$ is also inside the center of $\mathfrak{a}^\prime$, then the extension is called \emph{central}.

Consider the quotient $\mathfrak{n} \equiv \sigma \widetilde{\mathfrak{g}} / \sigma^2 \widetilde{\mathfrak{g}} = \sigma \mathfrak{g}_1$.
Obviously, $\mathfrak{n}$
is an ideal of $\mathfrak{g}_1$. Moreover, $\mathfrak{n}$ is an abelian ideal and is isomorphic\footnote{A Lie algebra $g_{\rm abelian}$ is isomorphic to the algebra $g$ as a vector space but all its commutators are vanishing.}
to $\left( \mathfrak{g}_0 \right)_{\rm abelian}$. Our first level extension is actually given by the following exact sequence:
\be
\label{exact}
0 \rightarrow \mathfrak{n} \rightarrow \mathfrak{g}_1 \xrightarrow{\pi}  \mathfrak{g}_0 \rightarrow  0  \, .
\ee
Since $\mathfrak{n}$ is by construction abelian, our $\mathfrak{g}_1$ extension is \emph{always} abelian.
The ideal $\mathfrak{n}$ is, however, not necessarily inside the center of $\mathfrak{g}_1$.
In fact, $\mathfrak{n}$  consists of elements of the form
$\sigma a_i$ and becomes central if and only if $\mathfrak{g}_0$ is abelian. This was exactly the case for the IW bundle defined by
(\ref{SO3scaling}), which led to the Weyl algebra (\ref{WeylJ}).

It seems, therefore, that our construction produces central extensions only upon very restricting conditions.
In particular, it rules out the Bargmann (non-zero mass) extension of the Galilean algebra.
Luckily, we can slightly modify the extension  described by (\ref{exact}).

Assume that $\mathfrak{m}_0$ is an ideal of $\mathfrak{g}_0$. It automatically implies that $\pi^{-1} \left( \mathfrak{m}_0 \right)$ is an ideal of $\mathfrak{g}_1$, where $\pi$ is the projection map from (\ref{exact}). Furthermore,
$\mathfrak{m}_1 \equiv \sigma \pi^{-1} \left( \mathfrak{m}_0 \right)$ is an abelian ideal of $\mathfrak{g}_1$ and the quotient
$\mathfrak{n}/\mathfrak{m}_1 = \sigma \mathfrak{g}_1/\mathfrak{m}_1$
is isomorphic to $\left( \mathfrak{g}_0/\mathfrak{m}_0 \right)_\textrm{abelian}$.
We get the following \emph{abelian} extension of $\mathfrak{g}_0$:
\be
\label{exact2}
0 \rightarrow \mathfrak{n}/\mathfrak{m}_1
                  \rightarrow \mathfrak{g}_1/\mathfrak{m}_1 \xrightarrow{\pi}  \mathfrak{g}_0 \rightarrow  0   \, .
\ee
The above extension is central if and only if $\mathfrak{g}_0/\mathfrak{m}_0$ is abelian.
Indeed, the latter requirement is equivalent to the statement that
$[ \mathfrak{g}_0, \mathfrak{g}_0] \subset \mathfrak{m}_0$. But that means that
$[ \mathfrak{g}_1, \mathfrak{g}_1] \subset \pi^{-1} \left( \mathfrak{m}_0 \right)$
and so $[ \sigma \mathfrak{g}_1, \mathfrak{g}_1] \subset \mathfrak{m}_1$. The latter immediately implies
that $\mathfrak{n}/\mathfrak{m}_1$ is central.

This way the number of the new generators in the extended algebra will be smaller than total number of generators in $g$.
We will see in the next section that (\ref{exact2}) naturally leads to the central (mass) extension of the Galilean algebra.

\section{\bf Examples}

We have already seen how the central extension of the Weyl algebra emerges from the $SO(3)$ contraction.
In the rest of the paper we will see other applications of the proposed extension method.

\subsection{\bf The non-central extensions of the Poincar\'e algebra from the de Sitter
algebra contraction}

In $d$ space-time dimensions the commutators of the de Sitter and the anti de Sitter algebras ($SO(d,1)$ and $SO(d-1,2)$) are:
\bea
\label{deSitter}
\left[ M_{\mu \nu}, M_{\lambda \rho} \right] &=& \eta_{\mu \lambda} M_{\nu \rho} + \eta_{\nu \rho} M_{\mu \lambda} -
                      \left( \lambda \leftrightarrow \rho \right)   \, , \nonumber \\
\left[ M_{\lambda \rho}, P_{\mu} \right] &=& \eta_{\mu \lambda} P_{\rho} + \eta_{\mu \rho} P_{\lambda} \, , \nonumber \\
\left[ P_{\mu}, P_{\nu} \right] &=& \varepsilon M_{\mu \nu}  \, .
\eea
Here $\eta = \textrm{diag} \left(-1,1,\ldots,1 \right)$ and  $\varepsilon=1,-1$ for the de Sitter
and the anti de Sitter algebra respectively.

We are interested in the zero cosmological constant limit, $\Lambda \to 0$, of the (anti) de Sitter algebra.
As such the algebra (\ref{deSitter}) is $\Lambda$-independent, so we have to properly rescale the operators
introducing $\Lambda$ into the commutators. The new $\Lambda$-dependent algebra will be isomorphic to the original one
for any $\Lambda$ except for $\Lambda=0$, where it should reduce to the Poincar\'e algebra.
The right rescaling of the generators is, in fact, well known in the literature.
The metric on the (anti) de Sitter space can be written as:
\be
\label{metric}
\textrm{d} s^2_{\textrm{(a)dS}}
     = R^2 \cdot \frac{\eta_{\mu \nu} \textrm{d} y^{\mu} \textrm{d} y^{\nu}}{1 + \varepsilon y^2/4 } \, ,
\ee
where $y^2 \equiv \eta_{\mu \nu} y^{\mu} y^{\nu}$ and $R={\Lambda}^{-1/2}$ is the (anti) de Sitter radius.
In terms of these coordinates the isometry generators are:
\begin{align}
\label{MP}
& M_{\mu \nu} =
  \eta_{\nu \lambda} y^\lambda \frac{\partial}{\partial y^\mu} - \eta_{\mu \lambda} y^\lambda \frac{\partial}{\partial y^\nu} \, ,  \nonumber \\
& P_{\mu} = \left( \varepsilon - \frac{y^2}{4} \right) \frac{\partial}{\partial y^\mu} +
             \frac{1}{2} \eta_{\mu \lambda}  y^\lambda y^\rho \frac{\partial}{\partial y^\rho}  \, .
\end{align}
In the limit $\left( R, y^\mu \right) \to  \left( \infty, 0 \right)$ with $x^\mu = R \cdot y^\mu$
kept fixed, the metric (\ref{metric}) reduces to the Minkowski metric, while the  leading scaling behavior
of the generators (\ref{JBPH}) is:
\be
\label{PoincareRescaling}
M_{\mu \nu} \to M_{\mu \nu} \, , \,\, P_{\mu} \to R \cdot P_{\mu} \, .
\ee
It may be also interesting to see the limit in terms of the $\left( \xi^\mu, \xi^d \right)$ coordinates
used for the (anti) de Sitter space embedding in $\mathbb{R}^{d,1}$:
\be
\xi^\mu \xi_\mu + \varepsilon \left( \xi^d \right)^2 = \varepsilon R^2 \, .
\ee
These coordinates are related to $y^\mu$'s by:
\be
\frac{\xi^\mu}{R} = \frac{y^\mu}{1 + \varepsilon y^2/4} \, , \quad \frac{\xi^d}{R} = \frac{1 - \varepsilon y^2/4}{1 + \varepsilon y^2/4}  \, .
\ee
In the Poincar\'e limit this reduces to $\left( \xi^\mu, \xi^d \right)= \left( x^\mu, R  \right)$ and so, analogously to the $SO(3)$ example
in the first paragraph of Introduction, the limit corresponds to the zooming around the $\xi^d = R$
point of the (anti) de Sitter space.

With this rescaling we find the following IW bundle:
\bea
\left[ P^\sigma_{\mu}, P^\sigma_{\nu} \right] &=& \sigma \cdot \varepsilon M^\sigma_{\mu \nu}   \, ,
\eea
where the affine parameter $\sigma$ is just the cosmological constant
$\sigma = R^{-2} = \Lambda$ and we left out the commutators from the first two lines in
(\ref{deSitter}) since they remain completely $\sigma$-independent.

As expected, the ``classical'' $\mathfrak{g}_0$ algebra is precisely the Poincar\'e algebra:
\bea
\label{Poincare}
\left[ M^\0_{\mu \nu}, M^\0_{\lambda \rho} \right] &=& \eta_{\mu \lambda} M^\0_{\nu \rho} \quad + \textrm{permutations} \, , \nonumber \\
\left[ M^\0_{\lambda \rho}, P^\0_{\mu} \right] &=& \eta_{\mu \lambda} P^\0_{\rho} + \eta_{\mu \rho} P^\0_{\lambda} \, , \nonumber \\
\left[ P^\0_{\mu}, P^\0_{\nu} \right] &=& 0 \, .
\eea
The next level ``semi-classical'' algebra $\mathfrak{g}_1$ includes also the operators $M_{\mu \nu}^\1=\sigma M_{\mu \nu}$ and
$P_{\mu}^\1=\sigma P_{\mu}$. Notice, however, that the translations $P_{\mu}^\0$ form an ideal of the Poincar\'e algebra
$\mathfrak{g}_0$. We therefore can use the modified exact sequence (\ref{exact2}) with $\mathfrak{m}_0$ being the
subalgebra of translations $P_{\mu}^\0$, so that the extension will include
only the $M_{\mu \nu}^\1$ generators and not $P_{\mu}^\1$'s. Defining $Z_{\mu \nu} = \varepsilon M_{\mu \nu}^\1$ we arrive at the Maxwell algebra:
\bea
\label{Maxwell}
\left[ M^\0_{\mu \nu}, Z_{\lambda \rho} \right] &=& \eta_{\mu \lambda} Z_{\nu \rho} + \eta_{\nu \rho} Z_{\mu \lambda} -
                      \eta_{\mu \rho} Z_{\nu \lambda} - \eta_{\nu \lambda} Z_{\mu \rho}   \, , \nonumber \\
\left[ P^\0_{\mu}, P^\0_{\nu} \right] &=& Z_{\mu \nu} \, ,
\eea
with the first two commutators in (\ref{Poincare}) remaining intact. The extension is non-central since $Z_{\mu \nu}$
does not commute with the Lorentz generators. It becomes a scalar only for $d=2$.
Following the recipe from the previous section we can extend the algebra to higher levels.
The commutators between $Z_{\mu \nu}$'s become non-trivial already at the $\mathfrak{g}_2$ algebra:
\be
\label{ZY1}
\left[ Z_{\mu \nu}, Z_{\lambda \rho} \right] = \eta_{\mu \lambda} Y_{\nu \rho} + \eta_{\nu \rho} Y_{\mu \lambda} -
                      \eta_{\mu \rho} Y_{\nu \lambda} - \eta_{\nu \lambda} Y_{\mu \rho}   \, ,
\ee
where $Y_{\mu \nu} \equiv M^\2_{\mu \nu}=\sigma^2 M_{\mu\nu}$ and:
\bea
\label{ZY2}
\left[ P^\0_{\mu}, P^\1_{\nu} \right] &=& Y_{\mu \nu} \, .
\eea
For similar extensions of the Maxwell algebra recently studied in the literature
see \cite{Soroka:2006aj}, \cite{Bonanos:2008kr}, \cite{Bonanos:2008ez} and \cite{Gomis:2009dm}.

Before ending this section let us remark that the de Sitter algebra (\ref{deSitter}) permits an
additional, infinite cosmological constant limit \cite{Aldrovandi:1998ux}. For fixed $x^\mu = R \cdot y^\mu$ but with
$\left( R, y^\mu \right) \to  \left( 0,\infty\right)$ the momenta scale like $P_{\mu} \to P_{\mu}/R$. The contracted algebra
is equivalent to the Poincar\'e but its physical meaning is different, since the $P^\0_{\mu}$'s now are special conformal transformations
and not translations.

\subsection{\bf The Galilean contraction of the Poincar\'e algebra}

Our next example is the non-relativistic contraction of the Poincar\'e algebra.

In $d$ space-time dimensions the Poincar\'e algebra consists of\footnote{
We will use bold font to indicate $3$-dimensional space vectors.}
$(d-1)(d-2)/2$ space-space rotations $\mathbf{J}$,
$d-1$ boosts $\mathbf{B}$, $d-1$ space translations $\mathbf{P}$ and the time translation $H$.
In all of the remaining examples but the last one we will explicitly consider the $d=4$ case,
with a straightforward generalization for other dimensions.
The commutators are\footnote{Here $\left[ \mathbf{A}, \mathbf{B} \right] = \mathbf{C}$ is a shorthand notation for
$\left[ A_i, B_j \right] = \epsilon_{ijk} C_k$ and $\left[ \mathbf{A}, \mathbf{B} \right] = C$ stands for
$\left[ A_i, B_j \right] = \delta_{ij} C_k$. }:
\be
\label{PoincareJBPH1}
\left[ \mathbf{J}, H \right] = 0, \,\,\, \left[ \mathbf{J}, \mathbf{J} \right] = \mathbf{J}, \,\,\,
  \left[ \mathbf{J}, \mathbf{B} \right] = \mathbf{B}, \,\,\, \left[ \mathbf{J}, \mathbf{P} \right] = \mathbf{P}
\ee
and:
\be
\label{PoincareJBPH2}
\left[ H, \mathbf{B} \right] = \mathbf{P}, \quad \left[ \mathbf{B}, \mathbf{B} \right] = -\mathbf{J},
  \quad \left[ \mathbf{P}, \mathbf{B} \right] = H \, ,
\ee
with all other commutators vanishing.
The first set of the commutators
(\ref{PoincareJBPH1}) simply implies that $\mathbf{B}$, $\mathbf{P}$ and $\mathbf{J}$ transform as vectors under space rotations,
while $H$ is a scalar.
The differential representation of the operators in terms of the Minkowskian coordinates is:
\begin{align}
\label{JBPH}
& J_{i} = - \epsilon_{ij}^{\,\,\,\,k} x^j \frac{\partial}{\partial x^k}  \, ,  \quad
       B_{i} = \delta_{ij} x^j \frac{\partial}{\partial x^0} + x^0 \frac{\partial}{\partial x^i} \, , \nonumber \\
& \qquad   P_{i} = \frac{\partial}{\partial x^i}  \, , \qquad  H = \frac{\partial}{\partial x^0}  \, .
\end{align}
The non-relativistic limit corresponds to $c, x^0 \to \infty$ with $t=x^0/c$ held fixed.
The Poincar\'e algebra generators scale in this limit as:
\be
\label{GalileanRescaling}
\mathbf{J} \to \mathbf{J} \, , \,\, \mathbf{B} \to c \mathbf{B} \, , \,\,
\mathbf{P} \to \mathbf{P} \, , \,\, H \to \frac{H}{c} \, .
\ee
This rescaling leads to the following IW bundle:
\be
\label{GalileiIW}
\left[ H, \mathbf{B} \right] = \mathbf{P}, \quad \left[ \mathbf{B}, \mathbf{B} \right] = - \sigma \mathbf{J},
  \quad \left[ \mathbf{P}, \mathbf{B} \right] = \sigma H \, ,
\ee
where $\sigma = c^{-2}$ and for simplicity we dropped the $\sigma$-indices.  We also left out the $\sigma$-independent
part of the algebra (\ref{PoincareJBPH1}).
Putting $\sigma=0$ in (\ref{GalileiIW}) we of course find the Galilean algebra $\mathfrak{g}_0$:
\be
\label{Galilei}
\left[ H^\0, \mathbf{B}^\0 \right] = \mathbf{P}^\0, \,\, \left[ \mathbf{B}^\0, \mathbf{B}^\0 \right] = 0,
  \,\,  \left[ \mathbf{P}^\0, \mathbf{B}^\0 \right] = 0
\ee
together with the regular transformations of $H^\0$, $\mathbf{B}^\0$ and $\mathbf{P}^\0$ under the $\mathbf{J}^\0$ rotations.

As we have already emphasized before the $\mathfrak{g}_1$ extension has by construction twice more operators than $\mathfrak{g}_0$.
The new operators are $\mathbf{J}^\1$, $\mathbf{B}^\1$, $\mathbf{P}^\1$ and $H^\1$.
In order to get the central non-zero mass extension of (\ref{Galilei}) we have to use (\ref{exact2}), where the ideal
$\mathfrak{m}_0$ is the subalgebra of all Galilean rotations, boosts and space translations, namely $\mathbf{J}^\0$,
$\mathbf{B}^\0$ and $\mathbf{P}^\0$. This allows us to mode out $\mathbf{J}^\1$,
$\mathbf{B}^\1$ and $\mathbf{P}^\1$ from the extended algebra. Denoting $M=H^\1$ we find that $\mathfrak{g}_1/\mathfrak{m}_1$
is exactly the Bargmann algebra:
\be
\label{Bargmann}
\left[ H^\0, \mathbf{B}^\0 \right] = \mathbf{P}^\0, \,\, \left[ \mathbf{B}^\0, \mathbf{B}^\0 \right] = 0,
  \,\, \left[ \mathbf{P}^\0, \mathbf{B}^\0 \right] = M \, .
\ee
It is noteworthy that one can formally derive the Bargmann algebra by plugging $H=M/\sigma+H^\0$ into (\ref{GalileiIW}) and matching the
$\sigma$-expansion from the both sides. This approach is, however, not rigorously defined and unlike our method cannot be used for the next
order extension.

\subsection{\bf The Carrollian contraction of the Poincar\'e algebra}

The Carrollian contraction of the Poincar\'e algebra (\ref{PoincareJBPH1}, \ref{PoincareJBPH2})
is given by:
\be
\mathbf{J} \to \mathbf{J} \, , \,\, \mathbf{B} \to \frac{\mathbf{B}}{c} \, , \,\,
\mathbf{P} \to \mathbf{P} \, , \,\, H \to \frac{H}{c} \, .
\ee
It naturally follows from the $c \to 0$ limit of (\ref{JBPH}). The IW bundle is:
\be
\label{CarrollIW}
\left[ H, \mathbf{B} \right] = \sigma \mathbf{P}, \quad \left[ \mathbf{B}, \mathbf{B} \right] = - \sigma \mathbf{J},
  \quad \left[ \mathbf{P}, \mathbf{B} \right] = H \, ,
\ee
where this time $\sigma=c^2$. For $\sigma=0$ we find the Carrollian algebra:
\be
\label{Carroll}
\left[ H^\0, \mathbf{B}^\0 \right] = 0, \quad \left[ \mathbf{B}^\0, \mathbf{B}^\0 \right] = 0,
  \quad \left[ \mathbf{P}^\0, \mathbf{B}^\0 \right] = H^\0   \, .
\ee
By analogy with the Galilean case we can mode out some operators from the
next level $\mathfrak{g}_1$ extension. Notice that , $\mathbf{J}^\0$, $\mathbf{B}^\0$ and $H^\0$ form an ideal $\mathfrak{m}_0$
and so we can exclude $\mathbf{J}^\1$, $\mathbf{B}^\1$ and $H^\1$ from the $\mathfrak{g}_1$ extension.
With $\mathbf{Z} = \mathbf{P}^\1$ the $\mathfrak{g}_1/\mathfrak{m}_1$ algebra takes the following form:
\be
\left[ H^\0, \mathbf{B}^\0 \right] = \mathbf{Z}, \quad \left[ \mathbf{B}^\0, \mathbf{B}^\0 \right] = 0,
  \quad \left[ \mathbf{P}^\0, \mathbf{B}^\0 \right] = H^\0 \, .
\ee
This extension is central only for $d=2$ when $\mathbf{Z}$ becomes a scalar.

\subsection{\bf The exotic Newton-Hooke algebra}

In our discussion of the Galilean algebra extension we excluded the operator $\mathbf{J}^\1$ from the ``semi-classical'' algebra.
We did so because this operator is not central in the extended $4d$ Galilean algebra, for it does not commute with the $\mathbf{J}^\0$ rotations.
The situation changes for $d=3$, since in this case $\mathbf{J}$ is a scalar. The $d=3$ Galilean algebra with
the $J$ central extension is known in the literature as the exotic Galilean algebra.\footnote{One may also consider a similar
exotic central extension of the $3d$ Carrolllian algebra.}

Instead of focusing our attention on this algebra let us make the discussion a bit more general by starting with the Newton-Hooke (NH) contraction.
The NH algebra is the non-relativistic limit of the de Sitter algebra.
To be more precisely, one sends both $c$ and $R$ to infinity with the parameter $\omega=c/R$ held fixed.

For $d=3$ the de Sitter algebra is:
\be
\label{d3deSitter1}
        \left[ J, H \right] = 0 \, ,
  \quad \left[ J, B_i \right] = \epsilon_{ij} B_j \, ,
  \quad \left[ J, P_i \right] = \epsilon_{ij} P_j
\ee
and:
\begin{align}
&	    	\left[ H, B_i \right] = P_i \, ,
    \quad  \left[ B_i, B_j \right] = -\epsilon_{ij} J\, ,
    \quad \left[ P_i, B_j \right] = \delta_{ij} H   \, ,\nonumber \\
&  	\qquad \left[ P_i, H \right] = B_i \, ,
    \quad  \left[ P_i, P_j \right] = \epsilon_{ij} J  \, ,
\end{align}
where $(H,P_i)=P_\mu$, $B_{i}=M_{0i}$ and $J=M_{12}$.
Combining the rescalings (\ref{PoincareRescaling}) and (\ref{GalileanRescaling}):
\be
J \to J \, , \,\, B_i \to c B_i \, , \,\,
P_i \to \frac{P_i}{R} \, , \,\, H \to \frac{H}{c R}
\ee
we obtain a new IW bundle with the $\sigma=c^{-2}$ affine parameter:
\begin{align}
&	    	\left[ H, B_i \right] = P_i \, ,
    \,\,    \left[ B_i, B_j \right] = - \sigma \cdot\epsilon_{ij} J\, ,
    \,\,    \left[ P_i, B_j \right] = \sigma \cdot \delta_{ij} H   \, ,\nonumber \\
&  	\qquad \left[ P_i, H \right] = \omega^2 B_i \, ,
    \,\, \left[ P_i, P_j \right] = \sigma \cdot \epsilon_{ij} \omega^2 J  \, ,
\end{align}
where we omitted again all the commutators with $J$, because they are $\sigma$-independent.

Obviously the boosts $B^\0_{i}$ and the momenta $P^\0_{i}$ will form an ideal of $\mathfrak{g}_0$. Using this ideal for
(\ref{exact2}) we arrive at the \emph{exotic} $3d$ Newton-Hooke algebra:
\begin{align}
\label{eNH}
&	    	\left[ H^\0, B^\0_i \right] = P_i \, ,
    \,\,    \left[ B^\0_i, B^\0_j \right] = - \epsilon_{ij} \widetilde{Z} \, , \nonumber \\
&   \qquad \qquad    \left[ P^\0_i, B^\0_j \right] = \delta_{ij} M   \, ,\nonumber \\
&  	\left[ P^\0_i, H^\0 \right] = \omega^2 B^\0_i \, ,
    \,\, \left[ P^\0_i, P^\0_j \right] = \epsilon_{ij} \omega^2 \widetilde{Z}  \, ,
\end{align}
where $\widetilde{Z} \equiv J^\1$ and $M \equiv H^\1$. For $\omega=0$ (\ref{eNH}) transforms into
the exotic Galilean algebra.


\section{\bf Open questions}

In our paper we have shown that given an \.{I}n\"on\"u-Wigner contraction of one algebra to another, one can easily find an infinite
extension of the latter. The method works for both central and non-central extensions and the extension may be truncated
at any level.

There are plenty of open problems to be explored. Let us list some of them:
\begin{itemize}
	\item It is will be interesting to establish a connection between our extension approach and the expansion method presented in
		\cite{deAzcarraga:2002xi}. This method is somewhat similar in spirit to ours.
		It is based on the Maurer-Cartan one-forms expansion in powers of a real parameter which
	  is related to the rescaling of the Lie group coordinates.
	  Our results should also be compared to those of \cite{Bonanos:2008ez}.
	\item It is not clear what is the physical meaning of the new (in)finite algebras. In some cases the answer is already known. For example,
	the Maxwell algebra (\ref{Maxwell}) corresponds to a charged relativistic particle moving in a constant electromagnetic field.
	The meaning of the second level extension (\ref{ZY1},\ref{ZY2}) is yet to be understood.
	\newline
	
	\item Finding  irreducible representations of the infinitely extended algebras and their truncated versions might be a challenging problem.
\end{itemize}

\begin{acknowledgments}


SK would like to thank Frank Ferrari for collaboration on the early stages of this long term project.
We are grateful to Jarah Evslin, Carlo Maccaferri, Chethan Krishnan, Glenn Barnich, Laurent Houart,
Ricardo Argurio, Marc Henneaux, Jacob Sonnenschein, Anatoly Dymarsky, Ayan Mukhopadhyay and especially Daniel and Melvin
Persson for fruitful discussions.

The work of SK is supported by the European Commission Marie Curie Fellowship under the
contract IEF-2008-237488.

\end{acknowledgments}

\begingroup\raggedright\endgroup

\end{document}